\documentclass[superscriptaddress,floatfix,twocolumn,amsmath,amssymb]{revtex4-1}
\usepackage{tabu}
\usepackage{gensymb}
\usepackage{graphicx}
\usepackage{microtype}

\usepackage{graphicx}
\usepackage{dcolumn}
\usepackage{bm}
\usepackage{amssymb}
\usepackage{natbib}
\usepackage{mathtools}
\usepackage{color}
\usepackage{datetime}
\usepackage{footnote}
\usepackage{bbold}
\usepackage[normalem]{ulem}
\usepackage{dcolumn}
 \usepackage{ragged2e}
 \usepackage{xfrac}
 \usepackage{sidecap}
 \usepackage{setspace}
 \usepackage{multirow}
\usepackage{xcolor}
\usepackage{hyperref}
\usepackage{cleveref}
\usepackage{siunitx}
\usepackage[utf8]{inputenc}

\begin{document}

\title{Nanostructuring of LNOI for efficient edge coupling}
\author{Inna Krasnokutska}
\affiliation{Quantum Photonics Laboratory and Centre for Quantum Computation and Communication Technology, School of Engineering, RMIT University, Melbourne, Victoria 3000, Australia}

\author{Jean-Luc J. Tambasco}
\affiliation{Quantum Photonics Laboratory and Centre for Quantum Computation and Communication Technology, School of Engineering, RMIT University, Melbourne, Victoria 3000, Australia}

\author{Alberto Peruzzo}
\thanks{alberto.peruzzo@rmit.edu.au}
\affiliation{Quantum Photonics Laboratory and Centre for Quantum Computation and Communication Technology, School of Engineering, RMIT University, Melbourne, Victoria 3000, Australia}

\begin{abstract}
We present the design, fabrication and characterization of LNOI fiber-to-chip inverse tapers for efficient edge coupling. The etching characteristics of various LNOI crystal cuts are investigated for the realization of butt-coupling devices. We experimentally demonstrate that the crystal cut limits the performance of mode matching tapers. We report a butt-coupling loss of 2.5\,dB/facet and 6\,dB/facet by implementing 200\,nm tip mode matching tapers in $+Z$-cut LNOI and $X$-cut MgO:LNOI waveguides with low propagation loss.  We anticipate that these results will provide insight into the nanostructuring of LNOI and into the further development of efficient butt-coupling in this platform.
\end{abstract}


\maketitle

\section{Introduction} 

Lithium niobate on insulator (LNOI) and magnesium oxide doped LNOI (MgO:LNOI) have recently emerged as promising photonic platforms.  Many key components have been demonstrated in LNOI including low-loss waveguides \cite{Zhang:17}, electro-optical modulators \cite{Mercante:18, WangNat:18}, and wavelength converters  \cite{He:18}, while MgO:LNOI has been minimally investigated despite its superior high power properties over LNOI  \cite{Gunter1988, Sun:12, Wangmg:18}.  Efficient coupling of light into LNOI waveguides is a major challenge yet to be overcome before this platform can be competitive with high-index contrast platforms including silicon on insulator and silicon nitride \cite{Baets, Orobtchouk:00, Maire:08}. Broadband, efficient and polarization insensitive coupling of light into photonic chips is essential for devices including Mach-Zehnder modulators and wavelength converters, as well as for reliable packaging of photonic chips \cite{Sacher:14, Tremblay2017LargeBS, Liao2005HighSS}.

Several approaches including the use of grating couplers, lensed fibers, high numerical aperture fibers and inverse tapers have been used to improve coupling efficiency \cite{Baets, Papes:16, Maegami:15, 6895281}. Grating couplers, which are commonly used for vertical coupling into the chip, allow good coupling efficiency and the ability to access circuits inside a chip with alignment tolerances; however, compared to butt-coupling, limit the wavelength and polarization performance of devices and are prone to fabrication errors \cite{Baets,Chen:17,Jian:18}. Butt-coupling is insensitive in wavelength and polarization; however, suffers from high mode-mismatch losses unless mitigated, for example, with spot size converters via inverse tapering \cite{6895281, Maegami:15, Fu:14, PU20103678}.

It has been demonstrated that LNOI can be nanostructured \cite{Zhang:17, Mercante:18, Krasnokutska:18, Chen:17}, though the fabrication of small features required for inverse tapers is challenging due to the highly isotropic etching of LN in argon (Ar) plasma, dependent on crystal cut and MgO doping. The non-vertical sidewall angle limits the minimum feature size down to which LNOI may be structured.

In this paper, we analyze the material limitations in nanostructuring various LNOI and MgO:LNOI faces. We then design, fabricate and characterize mode-matching tapers on $+Z$-cut LNOI and $X$-cut MgO:LNOI with coupling loss of 2.5\,dB/facet and 6\,dB/facet respectively with a 200\,nm tip.

\section{Nanostructuring LNOI}
\begin{figure*}[t]
\centering
\includegraphics[width=.8\linewidth]{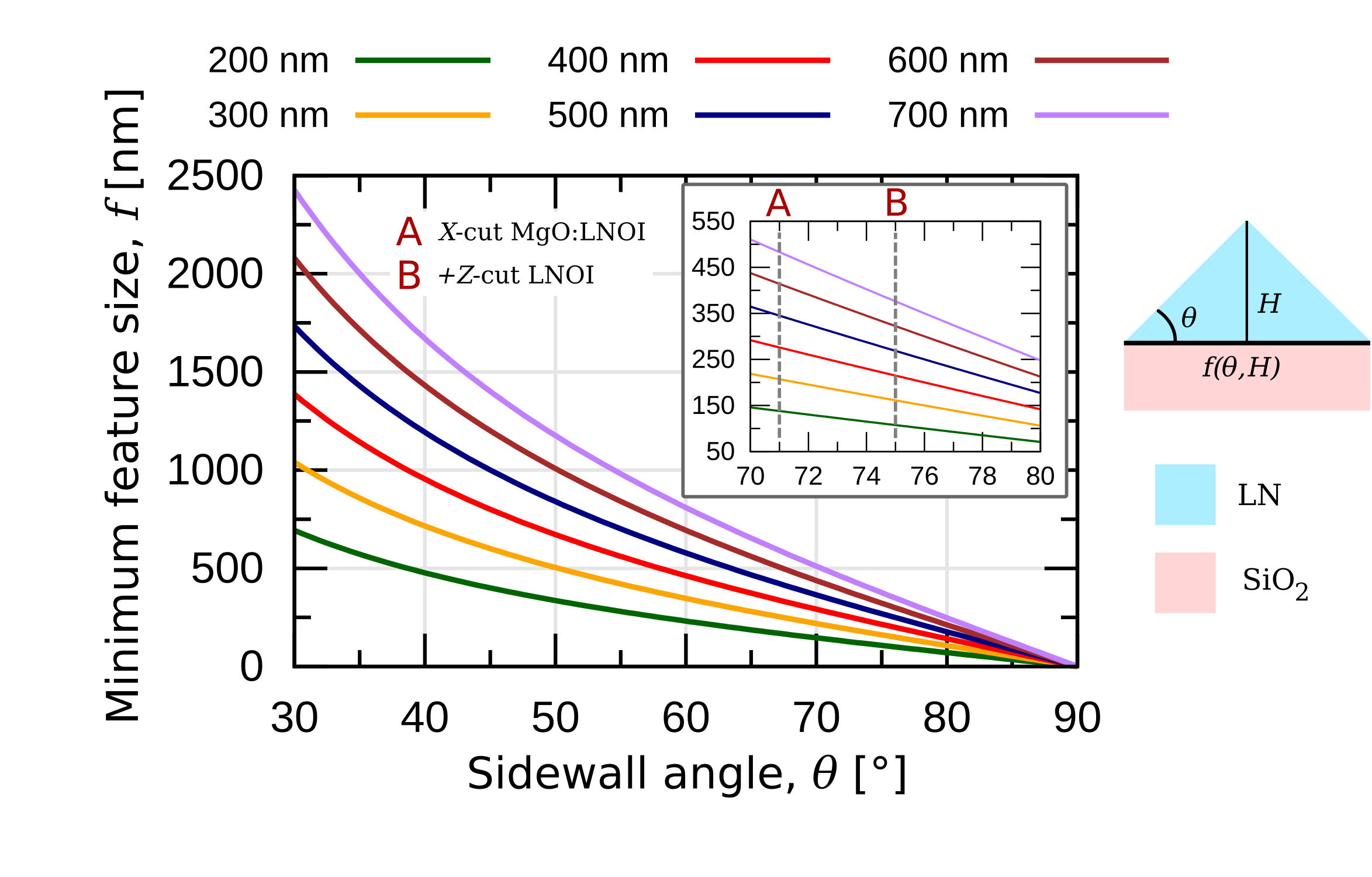}
\caption{Calculated minimum feature size versus sidewall angle for several etch heights. The minimum feature size that can be achieved for $+Z$-cut LNOI and $X$-cut MgO:LNOI are shown as inset. The schematic representation of minimum feature size definition is shown on the side of the picture. A thin LN film (blue) is seated atop SiO$_2$ (pink), where $f$ is the width of the minimum achievable feature and is dependent on the sidewall angle, $\theta$, and the etch depth, $H$.}
\label{fig1}
\end{figure*}

Recently, we demonstrated that by mixing argon and fluorine ions during plasma reactive ion etching (RIE) it is possible to achieve low loss rib waveguides with sidewall angle of $75\degree$ on $+Z$-cut LNOI \cite{Krasnokutska:18}. In this paper, we apply the developed fabrication process to a range of commercially available LNOI types, reporting their etching characteristics in Tab. \ref{tab:LNOIcompaarison}. The nanostructuring characterization enables to determine the minimum achievable feature size afforded by the particular cut and MgO doping of the LNOI platform. Figure \ref{fig1} shows the dependence of minimum feature size on sidewall angle. We calculate the the minimum bottom width of a feature that can be fabricated for different etching depths up to 700\,nm. It can be seen that sidewall angle strongly limits nanostucturing of LN films.
\begin{center}
\begin{table*}[t]
\caption{\label{tab:LNOIcompaarison} Etching characteristics of different LNOI types.}
\begin{tabular}{ m{2.5cm} | m{2cm} | m{2cm} | m{2cm} | m{2.5cm}} 
\hline
Material & Etching rate [nm/min] & Degree of Anisotropy, $D$ & Sidewall angle [$^{\circ}$] & Propagation loss [dB/cm] \\ 
\hline
LNOI $Z$-cut ($+Z$) & 18 & 0.86 & 75 & less than 0.1 \\ 
\hline
LN $Z$-cut ($-Z$) & 18.6 & -- & 78 & not on insulator\\
\hline
LNOI $X$-cut & 14.6 & 0.76 & 70 & less than 0.1  \\ 
\hline
MgO:LNOI $Z$-cut & 14 & 0.25 & 60 & 1 \\
\hline
MgO:LNOI $X$-cut & 16.6 & 0.45 & 71 & less than 0.1\\
\hline
\end{tabular}
\end{table*}
\end{center}
For example, the 50$\degree$ angles resulting from Ar milling will, in the best case, allow 500\,nm wide features etched 300\,nm deep, making the design and fabrication of efficient mode matching inverse tapers and gratings challenging.  On the other hand, $+Z$-cut LNOI has a $75^\circ$ etch angle, which allows feature widths down to $\sim 180$\,nm for the same 300\,nm etch depth.

In order to investigate limitations of the fabrication process, we choose five LN samples: $+Z$-cut LNOI and $X$-cut LNOI with film thickness 500\,nm, $+Z$-cut MgO:LNOI and $X$-cut MgO:LNOI with LN thin film thickness of 600\,nm, and $-Z$ LN bulk substrate 500\,$\mu$m thick. All samples are then patterned according to the process discussed in the \cite{Krasnokutska:18} to achieve a $\sim$1\,$\mu$m wide metal mask for rib waveguides. The samples are then RIE etched with the same recipe and their etching characteristics are deduced based on SEM imaging of the waveguide cross section as well as atomic force microscopy (AFM) measurements. Finally, the waveguides fabricated on LNOI and MgO:LNOI are cladded with 3\,$\mu$m of plasma enhanced chemical vapor deposition (PECVD) SiO$_2$, diced on a Disco DAD-321 using optical grade dicing, and their optical propagation losses measured. The propagation loss of $-Z$ LN bulk was not measured due to the absence of bottom cladding.

\begin{figure*}[t]
\centering
\includegraphics[width=.8\linewidth]{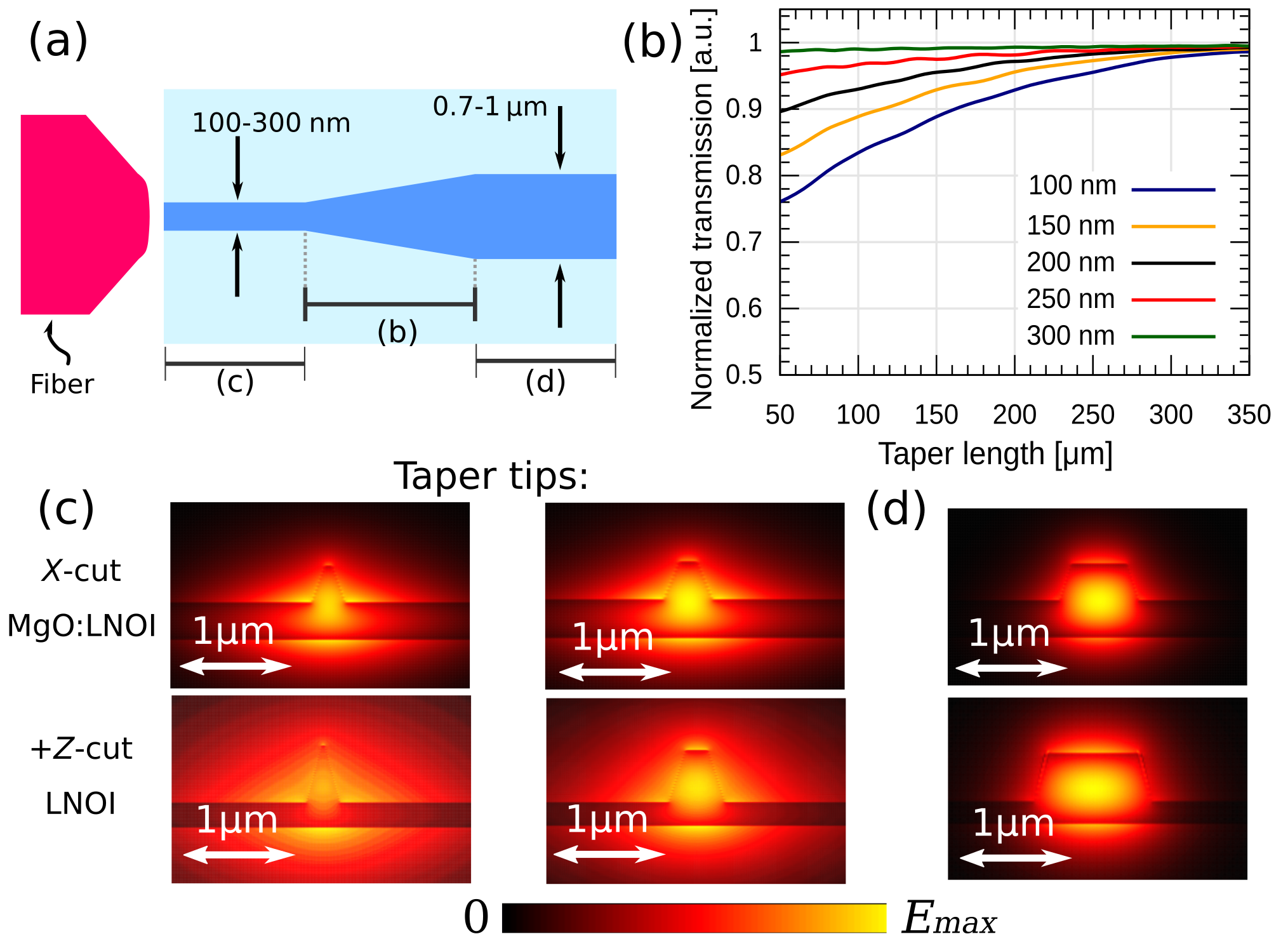}
\caption{Design of inverse taper: (a) schematic representations of light coupling into the waveguide via an inverse taper; (b) simulated dependence of the overall transmission through waveguide via taper length for different taper dimensions; (c) simulated mode profiles for tapers with different base widths for $X$-cut MgO:LNOI and $Z$-cut LNOI illustrating the increase in MFD;  (d) the simulated mode profiles for the untapered waveguides in $X$-cut MgO:LNOI and $Z$-cut LNOI discussed in this paper.}
\label{fig2}
\end{figure*}

We analyze various etching properties of LNOI including etch rate, sidewall angle, degree of anisotropy and propagation loss.  The degree of anisotropy, $D$, is defined as $D=1-B/2H$, where $B=W_{mask}$-$W_{final}$ is the etch bias, defining the amount of lateral etching, $W_{mask}$ and $W_{final}$ are the pre-etching and post-etching widths, and $H$ is the etch depth.

$Z$-cut LN is more chemically active in fluorine plasma than $X$-cut, as a result, it is possible to achieve faster etching rates, better anisotropy and near-vertical sidewall angles with $Z$-cut; this is reflected in our tests---we observe slower etching rates and a smaller degree of anisotropy for $X$-cut LNOI and we found that it possesses more shallow sidewall angle, however we did not observe effect of this on the optical propagation loss Tab. \ref{tab:LNOIcompaarison}, measured by using the Fabry-Perot technique and discussed in our previous papers \cite{Krasnokutska:18, krasnokutska2018large}. It is well known that $\pm X$ faces of LN have similar etching characteristics, meanwhile large differences can be observed for $+Z$ and $-Z$ faces, that was confirmed in our experiments and reported in the Tab. \ref{tab:LNOIcompaarison}. It was also shown that MgO:LNOI possesses different characteristics to not doped LNOI. It is more resistant to the etching process, moreover exhibits less anisotropy and $Z$-cut MgO:LNOI possesses the slowest etching rate and the most shallow angle compared to the other investigated LN crystal cuts. These results can be explained by its faster etching rate in lateral direction, which reduces the waveguide width resulting in increased propagation loss due to the enhanced light interaction with the waveguide sidewalls causing increased scattering losses.

\begin{figure*}[t]
\centering
\includegraphics[width=.8\linewidth]{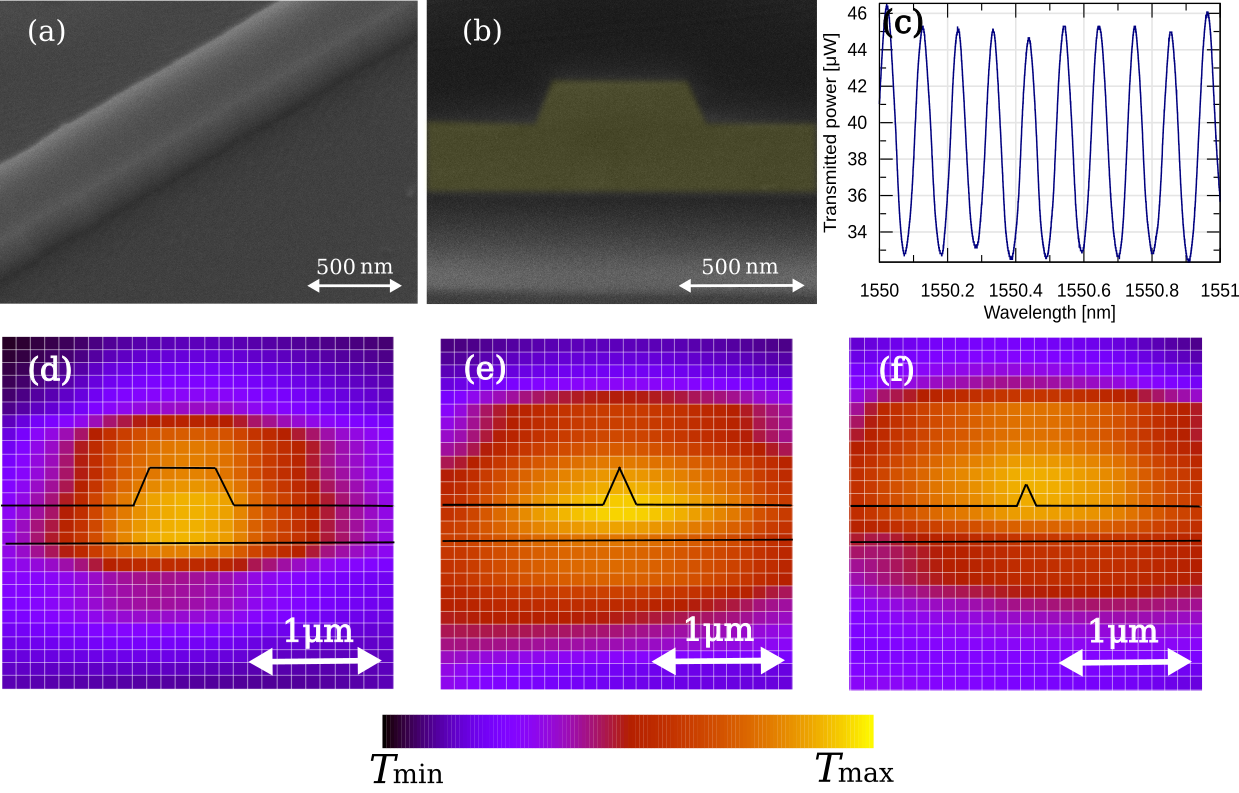}
\caption{Scanning electron microscope pictures of an etched single mode MgO:LNOI waveguide: (a) waveguide sidewall imaged at a $40^{\circ}$ tilt; (b) cross-section taken using FIB slicing and SEM imaging; the yellow false coloring highlights the waveguide outline. Optical characterization of fabricated MgO:LNOI mode-matching tapers and single mode waveguide: (c) Fabry-Perot measurements of propagation loss performed on 200\,nm inverse taper (the input laser power is 0.5\,mW); (d) measured optical power distribution at the output of the untapered waveguide; (e) measured optical power distribution at the output of 200\,nm base width taper and (f) $<100$\,nm base width taper illustrating the increase in MFD, where black lines schematically show the fabricated waveguide dimensions.}
\label{fig3}
\end{figure*}

\section{LNOI inverse taper}

The typical base width of a C-band single mode LNOI waveguide is $<1.2$\,$\mu$m in Fig. \ref{fig2}(d), resulting in a mode field diameter (MFD) of $\sim$1.5\,$\mu$m; meanwhile, the typical MFD of an optical single mode (SM) fiber is $\sim$10\,$\mu$m at 1550\,nm wavelength light, resulting in a significant mode mismatch between the optical fiber and the waveguide, increasing the butt-coupling loss. By exchanging the SM fiber for a lensed fiber, which has a $\sim$2.5\,$\mu$m MFD across the C-band, improved mode matching can be achieved; however, there is still significant mode mismatch. By combining lensed fibers with inverse tapers, a drastic improvement to the waveguide MFD can be obtained, enabling good mode matching.  The inverse tapers also have the added benefit of matching the effective index of the waveguide to that of the fibre, minimizing back reflections.

Fig. \ref{fig2}(a) shows a schematic representation of the designed spot size converter. The device consists of a linear inverse taper on LNOI, which gradually becomes thinner towards the chip facets. The light is coupled in and out of the chip through the tapered regions using lensed fiber, and the width of the tip, typically 100--300\,nm, is designed to achieve a good overlap between the fiber mode and the waveguide mode; a narrower tip width leads to a larger MFD, as can be seen from the simulated mode profiles shown in Fig. \ref{fig2}(c). The nominal mode field profile of the LNOI waveguide discussed in this paper is shown in Fig \ref{fig2}(d). To adiabatically match the waveguide width at the taper tip to the nominal waveguide width, the taper length must be sufficiently long. The length of the taper depends on the tip width and the nominal waveguide width---for smaller tip widths, longer tapers are required.  Simulations of the inverse taper length are performed using the eigenmode expansion (EME) solver provided by the commercially available software Lumerical Mode, and are presented in Fig \ref{fig2}(b); it can be seen that for lengths greater than $250\,\mu$m, all tapers, regardless of tip width, have negligible insertion loss.

We target tip widths $\sim200$\,nm in $+Z$-cut LNOI and $X$-cut MgO:LNOI to ensure a sufficiently large MFD.  It is more challenging to increase the MFD in the thicker 600\,nm $X$-cut MgO:LNOI film than the 500\,nm for $+Z$-cut LNOI, so we expect increased coupling loss.  Based on the minimum feature size calculations presented in Fig. \ref{fig1}, we expect to be able to achieve taper tips down to 170\,nm in $+Z$-cut LNOI and 210\,nm for a 300\,nm etch in $+Z$-cut LNOI and $X$-cut MgO:LNOI respectively.  To compensate for etch bias, the taper tip widths are increased in the mask layout to achieve wider metal lift-off features that then etch down to approximately the target width.

We lifted off 400\,nm and 500\,nm taper tips on $X$-cut MgO:LNOI resulting in etched base width tips $<$100\,nm and 200\,nm.  The fabricated MgO:LNOI waveguides sidewall is shown on the Fig. \ref{fig3}(a) taken prior to SiO$_2$ cladding; the sidewall roughness is estimated to be $\sim$ 2\,nm RMS as reported in our previous work \cite{Krasnokutska:18} and the the etched waveguide profile is cross-sectioned using focused ion beam (FIB) milling and then SEM imaged at a $50^{\circ}$ angle as shown in Fig. \ref{fig3}(b). The transmission spectrum of the 200\,nm taper is shown in the Fig. \ref{fig3}(c) corresponding to a propagation loss of $<0.1$\,dB/cm based on the Fabry-Perot technique. The laser light is coupled in and out of the chip via a pair of polarization maintaining (PM) lensed fibers. The output transmission of the waveguide is monitored over a $3 \times 3\,\mu$m window by sweeping the fiber position and recording the transmitted power at each point---this provides an indication of the waveguide MFD. The coupling loss of a nominal width (untapered) MgO:LNOI waveguide is 10\,dB/facet and the power distribution at the output is shown at the Fig. \ref{fig3}(d), meanwhile the 200\,nm taper achieves a transmission of 6\,dB/facet width the power distribution shown in Fig. \ref{fig3}(e). In general, a smaller taper tip leads to a larger waveguide MFD resulting in higher butt-coupling efficiency; however, the overall transmission through the $<$100\,nm taper drops as the tip becomes increasingly narrow due to the etch bias.  Towards the tip of the inverse taper, the MgO:LNOI film has excessively thinned and does not support a guided mode.  Figure \ref{fig3}(f) shows that butt-coupling is still possible as the inverse taper eventually becomes wide enough to pick up the leaky mode of the $<$100\,nm tip; however, the structure exhibits a high loss of $\sim$13\,dB/facet.

\begin{figure*}[t]
\centering
\includegraphics[width=0.8\linewidth]{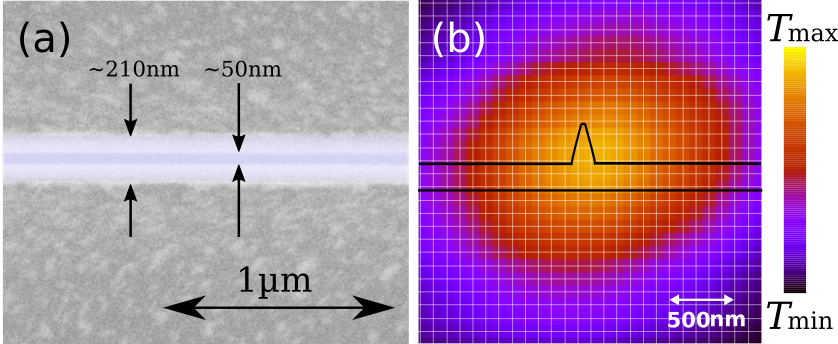}
\caption{(a) Scanning electron microscope of an etched taper in $Z$-cut LNOI tip; the false blue-coloring highlights the base and top edges of the taper tip. (b) Measured optical power distribution at the output of 200\,nm taper, where black lines schematically show the fabricated waveguide dimensions.}
\label{fig4}
\end{figure*}

Due to the significantly larger degree of anisotropy of the $+Z$-cut LNOI (over the $X$-cut MgO:LNOI), narrower taper tips of width 200\,nm and 300\,nm were lifted off and etched.  The 300\,nm lift-off mask resulted in the target $\sim$200\,nm etched tip width as shown in Fig. \ref{fig4}(a). We, again, record the power distribution at the taper output in Fig. \ref{fig4}(b) and measure an overall coupling loss of 2.5\,dB/facet.  The 200\,nm taper suffered similar film thinning issues as those seen with the narrower MgO:LNOI taper, resulting in a coupling loss of $\sim$12.5\,dB/facet.

The quality of the mode matching tapers is strongly affected by the etching properties of the (MgO:)LNOI. We observe improvements in coupling efficiency for both $X$-cut MgO:LNOI and $+Z$-cut LNOI at $\sim$200\,nm base taper tips widths. Due to the less anisotropic etching of $X$-cut MgO:LNOI, the performance of the spot size converters is strongly compromised.  Although etch bias may be compensated by increasing mask dimensions, the minimum feature size is limited by the etch angle and depth. 

\section{Conclusion}
The fiber-waveguide mode matching that can be achieved using the spot size converter is restricted by the etching properties of LN thin film. We have provided a detailed study of many fabrication limitations in LNOI platform. Even nanostructuring with a 75\,$^{\circ}$ waveguide sidewall angle, the minimum achievable feature size hinders the fabrication of $\sim$200\,nm taper tips needed for efficient butt-coupling. We demonstrated design and fabrication of LNOI inverse taper for improved mode matching between mode of an optical fiber and LNOI photonic components. We achieve butt-coupling efficiencies of 2.5\,dB/facet and 6\,dB/facet for $+Z$-cut LNOI and $X$-cut MgO:LNOI respectively, while preserving low propagation losses of <0.1\,dB/cm in both cases. These results play a critical role in the understanding of LNOI nanostructuring for photonics, and towards the development of efficient butt-coupling devices---a major obstacle for LNOI to become competitive photonics platform.

\section*{Funding}
Australian Research Council Centre for Quantum Computation and Communication Technology CE170100012; Australian Research Council Discovery Early Career Researcher Award, Project No. DE140101700; RMIT University Vice-Chancellors Senior Research Fellowship.

\section*{Acknowledgments}
This work was performed in part at the Melbourne Centre for Nanofabrication in the Victorian Node of the Australian National Fabrication Facility (ANFF) and the Nanolab at Swinburne University of Technology. The authors acknowledge the facilities, and the scientific and technical assistance, of the Australian Microscopy \& Microanalysis Research Facility at RMIT University.

\end{document}